\documentclass[aps,preprintnumbers,amsmath]{revtex4}
\usepackage{psfrag}
\usepackage{epsfig}
\usepackage{amsfonts} 
\usepackage{amssymb}  
\usepackage{graphicx} 
\usepackage{amsmath}  
\usepackage{amsmath}  
\usepackage{calligra}
\usepackage{mathrsfs}
\usepackage{tabularx}
\usepackage[latin1]{inputenc}
\usepackage{color}
\usepackage{suffix}
\usepackage{mathtools}
\usepackage{slashed}
\usepackage{ulem}
\usepackage{color}
\usepackage{suffix}
\usepackage{mathtools}
\usepackage{slashed}
\usepackage{ulem}
\usepackage{physics}
\usepackage{float}
\begin{document}

\title{Anomalous spontaneous induction of magnetic and electric fields in dense quark matter}

\author{E. J. Ferrer and J. M. Perez-Fernandez}
\affiliation{
Department of Physics and Astronomy, University of Texas Rio Grande Valley, 1201 West University Dr., Edinburg, TX 78539
}

\begin{abstract}

In this paper, we will demonstrate that a dense quark-matter system in the dual chiral density wave (DCDW) phase behaves as a ferromagnet in the sense that its magnetic-field dependent magnetization remains different from zero even at $B\rightarrow 0$. The corresponding permanent magnetization is a function of the baryonic chemical potential  $\mu$, decreasing up to zero as $\mu$ increases in the range of intermediate densities  ($312$ MeV $\leqslant \mu \leqslant 342$ MeV) and then increasing  from zero in the higher density interval $490$ MeV$\leqslant \mu \leqslant 550$ MeV. We will show that this system's ability to generate permanent magnetization, together with the existence of the axial anomaly, open up the possibility of spontaneously generating a magnetic field coupled to  a collinear electric field. The generated magnetic field can reach values up to $10^{16}$ G, depending on $\mu$, and the electric field will be 3 orders smaller. The fact that the DCDW phase is able to induce a magnetic field can be seen as its spontaneous tendency to remove the so called Landau-Peierls instability that is present in this single-modulated phase in the absence of a magnetic field. The spontaneous induction of a strong magnetic field at intermediate to high densities can be of interest for the astrophysics of compact stellar objects exhibiting strong magnetic fields as magnetars.

\end{abstract}


\date{\today}

\maketitle

\section{Introduction}

Quark matter at finite density has an interesting peculiarity. Once the Fermi sphere at zero temperature is created, the condensates associated with both the chiral pairs \cite{Buballa, Efrain-Review} and the color-superconducting Cooper pairs \cite{Casalbuoni} are energetically favorable when they are inhomogeneous. It has been found, on the other hand, that while the inhomogeneity of the Cooper pairs becomes favored when it has a higher-dimensional modulation, the inhomogeneous chiral condensate is favored when it is single modulated.
 However, single-modulated condensates suffer from the so-called Landau-Peierls (LP) instability \cite{Landau}. This instability appears  at any nonzero temperature, no matter how small. 
Due to this instability, the average of the order parameter vanishes, and the long-range order is washed out preserving only a quasi-long-range order \cite{Hidaka}, in close analogy with
smectic liquid crystals \cite{smectic}.

Nevertheless, when a magnetic field is present, the inhomogeneous chiral condensate phase known as the dual chiral density wave (DCDW) phase becomes energetically favored with respect to other known chiral phases  \cite{DCDW, Bo}, and moreover it  allows the formation of new structures in the
generalized Ginzburg-Landau (GL) expansion of the thermodynamic potential  \cite{LP-B} that remove the LP instability. That is, when considering the phonon fluctuations about the inhomogeneous condensate, these new field-induced structural terms give rise to a linear transverse mode in the spectrum of the fluctuations which prevents the existence of the LP instability \cite{LP-B, Will-2}. This is in sharp contrast  with what happens at $B=0$ in the DCDW phase, where the fluctuation spectrum is soft in the transverse direction and therefore, the system exhibits the LP instability. The new magnetized phase \cite{KlimenkoPRD82} satisfies an axion electrodynamics with anomalous electromagnetism and  is called the magnetic dual chiral density wave (MDCDW) phase \cite{Incera}.

We must emphasize that not all single modulated inhomogeneous phases in the presence of a magnetic field will be free from LP instability. In Ref. \cite{Will-2} it was shown that another necessary condition in removing the LP instability is the existence of a ground state that breaks time-reversal symmetry. This is the case for the ground state of the MDCDW phase, but not for other well-known single-modulated cases in the literature such as the real-kink crystal (RKC) \cite{Kink} and quarkyonic matter \cite{Quarkyonic} that can also be realized at finite density.

The phenomenology we are describing takes place in the range of moderate to high densities, that is, approximately between 1.8$n_s$ and 2.3$n_s$, or between 6.7$n_s$ and $10 n_s$ ($n_s$ is the saturation density), as can be obtained from the values of the baryonic chemical potential where the condensate amplitude $m$ and  modulation $b$ are not vanishing as can be seen in Fig. \ref{Fig. 1}.
These are densities that could well be found inside neutron stars (NS). Thus, the results we will report in this paper can be of interest to understand the physics of these compact objects taking into account, in addition, that 
 strong magnetic fields populate the vast majority of the astrophysical compact objects.  
 
 In this regard, the subclass of NS known as magnetars \cite{Magnetars} exhibits magnetic
fields that can reach values of the order of $10^{15}$ G on the star surface \cite{B-Surface}. Even magnetic fields of the order of
$10^{17}$ G and higher are expected to be sustained in stars interiors, as suggested by various recent observations (see, e.g., \cite{B-Observations} and
references therein), as well as from theoretical estimations.
Estimates based on the energy equipartition theorem, for instance, give upper estimates of the order of $10^{18}$ G for nuclear
matter \cite{Dong} and $10^{20}$ G for quark matter \cite{Portillo-2}. Nevertheless, fields of the order and higher than $10^{19}$ G create vanishing pressures that will make the star unstable \cite{Paulucci}.  
Other investigations of star stability taking into account the matter pressure and the star's gravitational
pressure in magneto-hydrostatic equilibrium point to fields of the order of $10^{17}$ G \cite{Moderate-Fields}. 

One mechanism widely considered to generate the star magnetic field is the so called 
 dynamo mechanism \cite{Dynamo-Mechanism}. This mechanism is responsible for the conversion of kinetic energy of an electrically
conducting fluid into magnetic energy. Nevertheless, when  this mechanism is considered taking
the star spin periods from the McGill Online Magnetar Catalog  \cite{McGill Catalog},
the magnetic-field strengths found are of the order of those observed in millisecond pulsars \cite{Millisecond-Pulsars},  which are far from the $10^{17}$ G inner field expected for magnetars \cite{Bo-Efrain}. 
 This result suggests that the origin of this field could be associated with a
microscopic mechanism that should be able to boost the inner field to larger values.

In this paper we will discuss how if the dense stellar medium is in the DCDW phase, it can behave as a ferromagnet that can generate, depending on the star inner density, 
magnetic fields much larger than those obtained by the dynamo mechanism in magnetars.

The paper is organized as follows. In Sec. 2, we express in a compact way the GL expansion formula for the thermodynamic potential introduced in Ref. \cite{Will}. Using this formulation, we study the magnetic field dependence of the minimum solutions of the condensate amplitude and modulation for different density values. In Sec. 3, we investigate the anomalous ferromagnetic properties of the MDCDW phase. We show that the magnetization as a function of the magnetic field tends as $B \rightarrow 0$ to a permanent magnetization that depends on the density. From Sec.'s 2 and 3 we see that both the minimum solutions for the condensate amplitude and modulation, as well as the magnetization have analytical behaviors in the  $B \rightarrow 0$ limit.  In Sec. 4, we show that in the DCDW phase of dense quark matter, the breaking of time-reversal and rotational symmetries by the ground state allows the induction of a magnetic field that can be as high as $10^{14} - 10^{16}$ G depending on the medium density. Taking into account the electromagnetic anomaly that is present in this phase, we show in Sec. 5 that together with the magnetic field a collinear electric field that is three orders smaller than $B$ can be induced. In Sec. 6, we present the concluding remarks and discuss how the findings we are reporting can affect the physics of NS.

\section{Amplitude and modulation of the inhomogeneous condensate of the  MDCDW phase versus magnetic field in the Ginzburg-Landau expansion}

It has been shown that at finite baryon density  the two condensates
\begin{equation}\label{DCDW-cond}
\langle \bar{ \psi } \psi \rangle = \Delta \cos( q_{ \mu } x^{ \mu } ), \qquad \langle \bar{ \psi } i \tau_{ 3 } \gamma_{ 5 } \psi \rangle = \Delta \sin( q_{ \mu } x^{ \mu } ) ,
\end{equation}
get expectation values different from zero forming a dual chiral density wave condensate \cite{DCDW}. In the presence of a magnetic field, 
the modulation vector is energetically favored when oriented along the field direction  \cite{KlimenkoPRD82, PLB743}. The phase with a ground state given by the inhomogeneous condensate (\ref{DCDW-cond}) in the presence of a magnetic field is called the MDCDW phase. The condensate with $\gamma^5$ matrix spontaneously breaks time-reversal symmetry. 

Under an external uniform magnetic field the system original global SO(3) symmetry decreases to SO(2) contributing to critical topological effects \cite{Incera, PLB743}. First of all, the magnetic field produces an asymmetry with respect to the zero energy level in the spectrum of the lowest
Landau level (LLL) \cite{KlimenkoPRD82}. 
\begin{equation}\label{LLL}
E_0=\epsilon\sqrt{m^2+p_3^2}+b, \quad l=0, \quad \epsilon = \pm,
\end{equation}
while the  highest Landau levels (HLL) are symmetric
\begin{equation}\label{HLL}
E_l=\epsilon\sqrt{\left(\varepsilon\sqrt{m^2+p_3^2}+b\right)^2+2|e_fB|l}, \quad l=1,2,3,...,
\end{equation}
Here, $m=-2G\Delta$ and $b=q_3/2$ are the condensate amplitude and modulation respectively.

The HLL spectrum ($l \geqslant 1$) has four branches, with $\varepsilon = \pm$ indicating spin projections and  $\epsilon = \pm$  the energy
sign. In contrast, the LLL ($l = 0$) has only two branches because only one spin projection contributes to the LLL modes. 
This asymmetry, consequently, gives rise to a topological term in the thermodynamic potential that significantly
enhances the window of inhomogeneity \cite{KlimenkoPRD82,PLB743}. Furthermore,
in the presence of an electric field with a nonzero component in the direction
of the background magnetic field, new topological effects emerge due to
the lack of invariance of the path-integral fermion measure under the local
chiral transformation. This lack of invariance gives rise to an ill-defined
Jacobian that requires a proper regularization \cite{Incera}. Using Fujikawa's method \cite{Fujikawa} it was extracted  in \cite{Incera} the regularized contribution to the effective action, which turned out to be the chiral anomaly in the electromagnetic sector $(\kappa /8)\theta F^{\ast}_{\mu \nu} F^{\mu \nu}$, with $\kappa/8=\alpha/4\pi$.
This interaction couples the electromagnetic strength tensor and its dual to $\theta = 2 b z $, with
$z$ the spatial coordinate in the direction of
the modulation. This chiral-anomalous term leads to anomalous electric
transport properties (for more details see the review \cite{Efrain-Review}). 

Considering that the condensate parameters $m$ and $b$ are smaller than the baryonic chemical potential $\mu$, it was carried out in  \cite{Will} the GL expansion of the thermodynamic potential in the weak-field approximation. To order $2N$ with $N = 1,2,3,\cdots$, it can be expressed as
\begin{equation} \label{omega}
\Omega^{ (2 N) } = \sum_{ i = 1, 2, 3, \cdots }^{ N } m^{ 2 i } \left[ \sum_{ j = 0, 1, 2, \cdots }^{ N - i } \alpha_{ 2 ( i + j ), 2 j } b^{ 2 j } + \sum_{ j = 0, 1, 2, \cdots }^{ N - ( i + 1 ) } \beta_{ 2 ( i + j ) + 1, 2 j + 1 } b^{ 2 j + 1 } \right],
\end{equation}
where the expansion coefficients $\alpha$ and $\beta$ depends at $T=0$ on the chemical potential and magnetic field. In \cite{Will} it was found a relationship that gives the option of finding all the coefficients as a numerical factor times the coefficient with the smallest subindices' values,
\begin{equation}\label{generic-coeff}
	c_{ n, n_{ q } } = \frac { ( n - 2 )! 2^{ 1 - \frac{ ( n - n_b ) }{ 2 } } } { ( \frac{ n - n_{ b } }{ 2 } )! n_{ b }! ( n - n_{ b } - 2 )!! }{ c_{ n, n - 2 } }, 
\end{equation}
here, $c$ is a generic coefficient representing $\alpha$ or $\beta$ without distinction and in our notation $n=2(i+j)$ when $c=\alpha$ and $n=2(i+j)+1$ when $c=\beta$, and $n_b$ denotes the exponent of $b$ in each case.  The smallest values in the subindices in (\ref{omega}) are for $j=0$ and $i=1$, which gives for the $\alpha$ coefficient,  $\alpha_{2,0}$; and for the $\beta$ coefficient, $\beta_{3,1}$.  The generic coefficients for the last member of each family (i.e. for the family formed by the coefficients with the same first index and with the higher second index) in the right-hand-side of \eqref{omega}, are given by
\begin{equation}\label{alpha}
	\alpha_{ 2 ( j + 1 ), 2 j } \approx \frac{ \delta_{ 0, 2 j } }{ 4 G } + \sum_{ r = 0, 2, 4, \cdots } \abs{ e B }^{ r } \frac{ B_{ r } }{ r ! } \cdot \frac{ 1 + 2^{ r } }{ 2 \pi^{ 2 } 3^{ r - 1 } } \cdot \frac{ 1 }{ ( 2 j - 1 )!! } I_{ 2 ( j + r - 1 ) }( \mu ),
\end{equation}

\begin{equation}\label{beta}
	\beta_{2(j+1) + 1, 2j+1} = - \frac{ 3 \abs{ e B } }{ (2 \pi )^{ 2 } }\frac{ 1 }{ (2j+1) \mu^{ (2j+1)} }, 
\end{equation}
where the $\beta$ coefficients exclusively depends on odd powers of $\mu$ and the $\alpha$ coefficients are described by the whole range of $j$ values. In (\ref{alpha}) the following notation was introduced 

\begin{equation}\label{I-1}
	I_{ - 2 }( \mu ) = - \frac{ 1 }{ 4 } \Lambda^{ 2 } + \frac{ 1 }{ 2 } \mu^{ 2 },
\end{equation}

\begin{equation}\label{I-2}
	I_{ 0 }( \mu ) = - \frac{ \gamma }{ 2 } - \ln( \frac{ 2 \mu }{ \Lambda } ),
\end{equation}

\begin{equation}\label{I-3}
	I_{ p > 0 }( \mu ) = - \frac{ 1 }{ p } \left( \frac{ i \sqrt{ 2 } }{ \Lambda } \right)^{ p } + \frac{ 1 }{ p!! }( p - 1 )! \mu^{ - p }.
\end{equation}
Here, we are considering the values for the theory cutoff $\Lambda$ and coupling constant $G$ as in \cite{KlimenkoPRD82}, given respectively by 
$\Lambda\approx 636.790$ MeV and $G\Lambda^2=6$. Also, we have that $ B_{ r }$ with $r=0,2,4,\cdots$ are the Bernoulli numbers and $\gamma$ is the Euler-Mascheroni constant ($\gamma\approx 0.577216$).

We would like to draw attention to the fact that the terms with beta coefficients in Eq. (\ref{omega}) depend only on linear powers in the field. This means that their contributions come only from the LLL. The terms with alpha coefficients have field powers greater than one, so they come from the HLL. This means that any anomalous effects that may occur will be associated with terms from the beta part of (\ref{omega}), which are associated with the asymmetric energy spectrum (\ref{LLL}).

 \begin{figure}
\begin{center}
\begin{tabular}{ccc}
  \includegraphics[width=9cm]{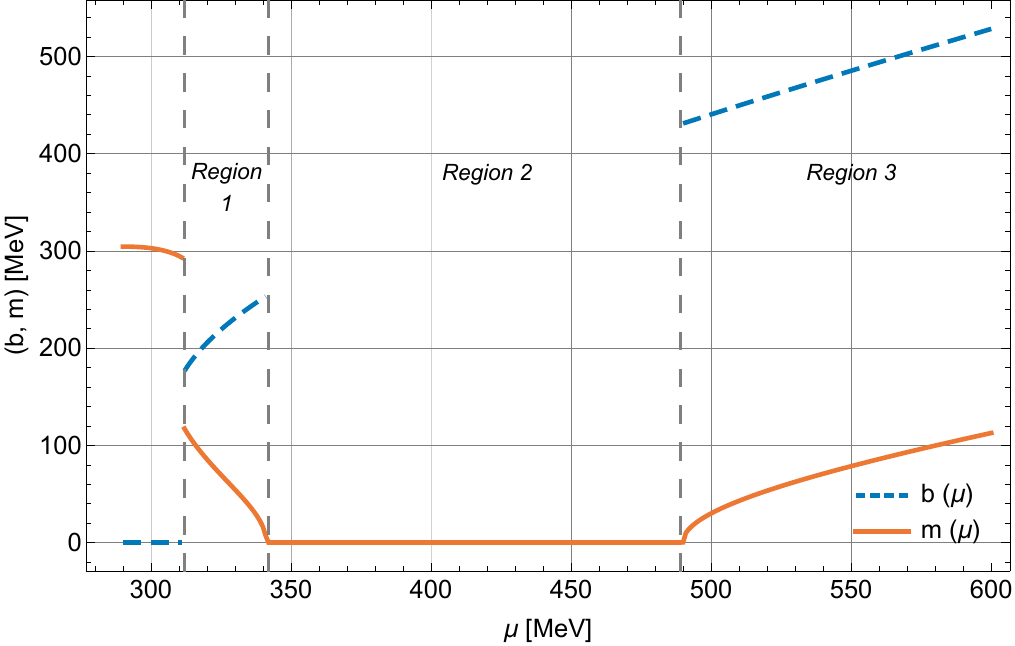} & \includegraphics[width=9cm]{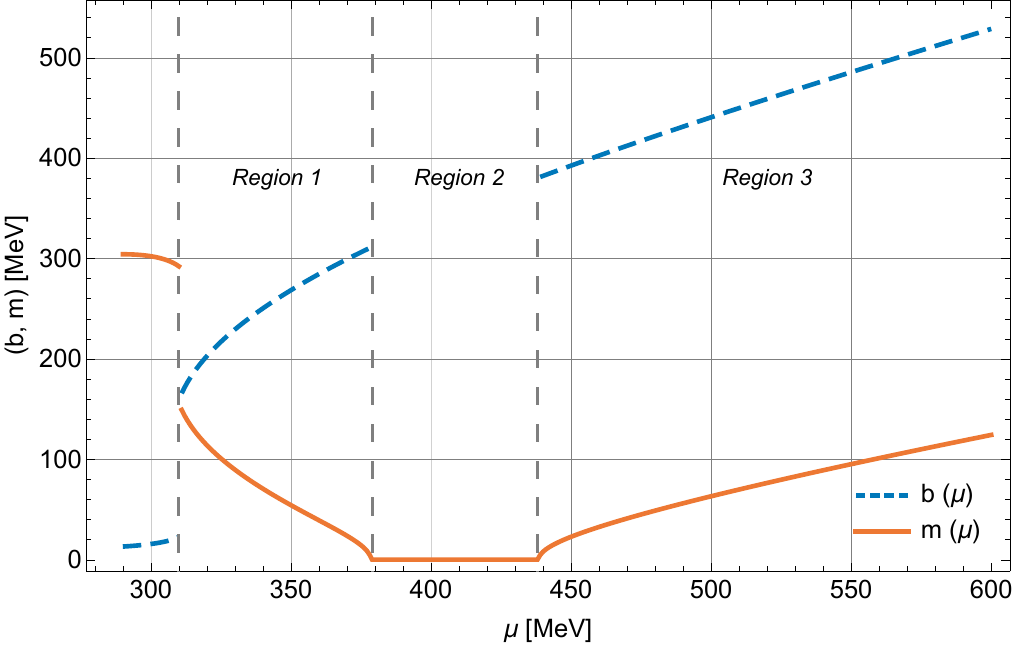}\\
(a) & (b)  \\
  \\
  \end{tabular}
    \end{center}
    \caption {(Color online) Minimum solutions at $T=0$ of the condensate amplitude $m$ and modulation $b$ as function of chemical potential, at $B=0$ in (a) and at $B=1.5 \times10^{18}$ G in (b).} 
     \label{Fig. 1}
\end{figure}

The minimum equations from where to obtain the condensate amplitude and modulation are respectively given by

\begin{eqnarray} \label{min-m}
\frac{ \partial \Omega^{ ( 2N ) } }{ \partial m } &=& \sum_{ i = 1 }^{ N } 2 i m^{ 2 i - 1 }  \left[ \sum_{ j = 0, 1, 2, \cdots }^{ N - i } \alpha_{ 2 ( i + j ), 2 j } b^{ 2 j } + \sum_{ j = 0, 1, 2, \cdots }^{ N - ( i + 1 ) } \beta_{ 2 ( i + j ) + 1, 2 j + 1 } b^{ 2 j + 1 } \right]=0
\end{eqnarray}

\begin{equation}\label{min-b}
\frac{ \partial \Omega^{(  2 N ) } }{ \partial b } = \sum_{ i = 1 }^{ N } m^{ 2 i } \left[ \sum_{ j = 0, 1, 2, \cdots }^{ N - i } 2 j \alpha_{ ( i + j ), 2 j } b^{ 2 j - 1 } + \sum_{ j = 0, 1, 2, \cdots }^{ N - ( i + 1 ) } ( 2 j + 1 ) \beta_{ 2 ( i + j ) + 1, 2 j + 1 } b^{ 2 j } \right]=0
\end{equation}

As a function of the chemical potential  the minimum solutions for $m$ and $b$ are given at $B=0$  in panel (a) of Fig.\ref{Fig. 1} and at $B=1.5 \times10^{18}$ G in panel (b).
From Fig.\ref {Fig. 1}-(a), it can be seen that in the interval $312$ MeV $\leqslant \mu \leqslant 342$ Mev the condensate amplitude $m$ decreases with $\mu$, while for $\mu \geqslant 490$ MeV it increases. At the same time, $b$ increases in all intervals where $m \neq 0$, stressing that the increase of the chemical potential favors the inhomogeneity.  This picture is indicating that at relatively small chemical potentials (i.e. for $312$ MeV $\leqslant \mu \leqslant 342$ Mev) the participation of particle-antiparticle pairs in the condensate is larger than that of particle-hole pairs, while at relatively larger chemical potentials (i.e. larger than $490$ MeV in this case) it will be the opposite \cite{Will-2}.
That is, this result is showing that increasing $\mu$, it becomes more difficult to pair the participating particles and antiparticles inside the condensate pairs, since they are separated by the Fermi energy (Notice that the Fermi sphere is formed starting from $\mu=312$ MeV, which is where the value of the chemical potential reaches the value of the mass at $\mu=0$); while, the particle-hole pairs formed in the proximity of the Fermi sphere will be reinforced,
so the condensate amplitude $m$ and modulation $b$ start to increase with $\mu$ at $\mu \geqslant 490$ MeV. The inhomogeneous condensate is reinforced in this case  as a consequence of the magnetic moment alignment due to the interaction with the magnetic field inside the particle-hole pairs \cite{Modulation}.

Hence, it can be concluded that depending on the internal composition of the pairs that contribute most to the condensate at a given density, the behavior of the parameters $m$ and $b$ with the chemical potential will be.
Thus, as indicated in Fig. \ref{Fig. 1}, Region 1 (limited to the interval  $312$ MeV $\leqslant \mu \leqslant 342$ MeV)  denotes where the greatest contribution is coming from particle-antiparticle pairs,  Region 2 (limited to the interval  $342$ MeV $\leqslant \mu \leqslant 490$ MeV), where there is equal participation of particle-antiparticle and particle-hole, and Region 3 (for $\mu > 490$ MeV) where there is greater participation of particle-hole. In Fig. \ref{Fig. 1} it is shown  how $m$ and $b$ behave with the variation of the baryonic chemical potential in each one of these regions.

\begin{figure}

    \begin{center}
        
        \begin{tabular}{ccc}
        
        \includegraphics[width=9cm]{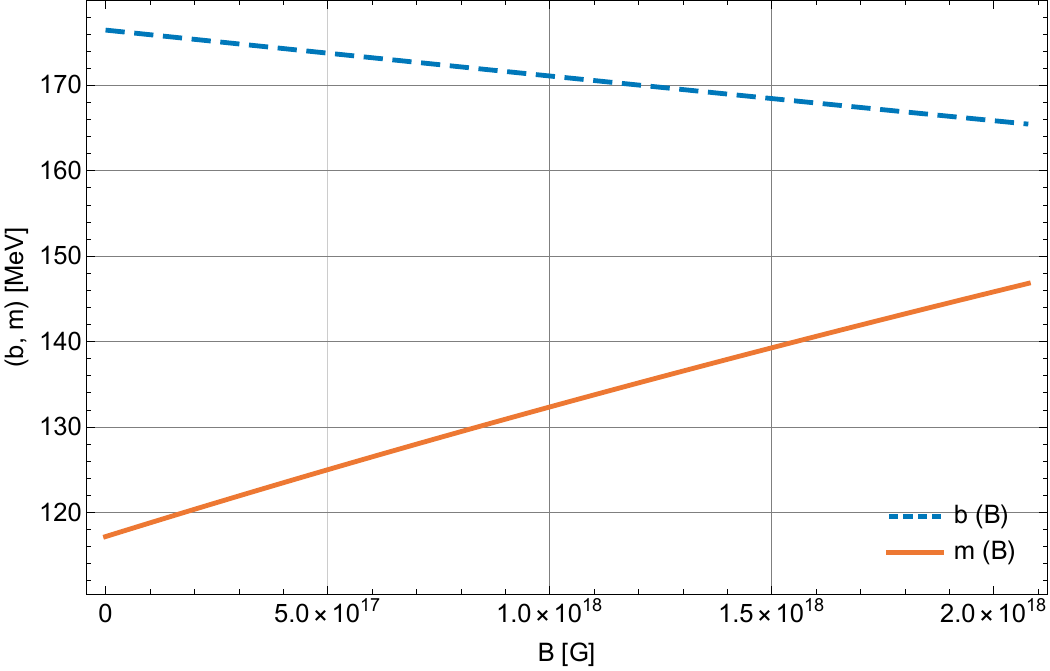} & \includegraphics[width=9cm]{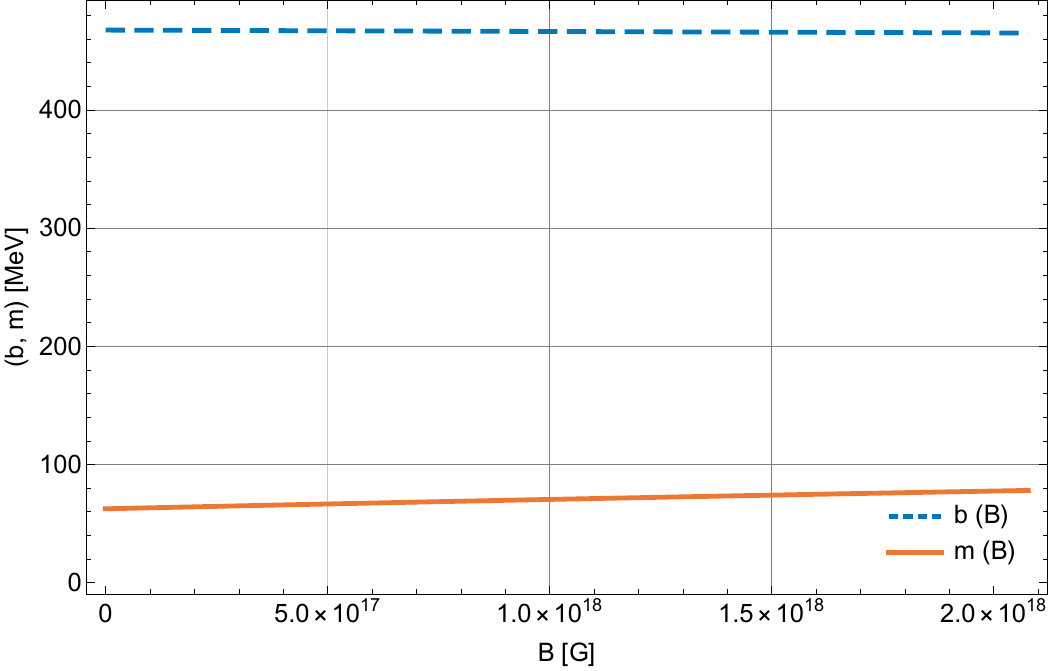} \\
        (a) & (b)  \\
        \\
        \end{tabular}
    \end{center}
    
     \caption {(Color online) Minimum solutions at $T=0$ of the condensate amplitude $m$ and modulation $b$ as function of the magnetic field , at $\mu=312$ MeV in (a) and at $\mu=530$ MeV in (b).}

        \label{Fig. 2}

\end{figure}

On the other hand, the role of a strong magnetic field is to shrink Region 2 as can be seen comparing Fig. \ref{Fig. 1}-(a) with Fig. \ref{Fig. 1}-(b).
Looking at Fig. \ref{Fig. 1}-(b), we see that the effect of the magnetic field is to shift the boundary between Region 1 and 2 to the right, meaning that the phase where the particle-antiparticle contribution is more relevant can now survive at larger chemical potentials. Also, from Fig. \ref{Fig. 2}-(a) we see that while the condensate amplitude $m$ increases with the magnetic field the modulation $b$ decreases. These results are compatible with the well known magnetic-field effect on particle-antiparticle pairs, characteristic of the magnetic catalysis of chiral symmetry breaking (MCCSB) phenomenology \cite{Magnetic Catalysis}. 
In the MCCSB at $\mu=0$ the magnetic field tends to strengthen the homogeneous condensate by producing a stronger alignment of the magnetic moments of the particle and antiparticles forming the homogeneous chiral condensate.  Thus, a magnetic field tends to favor oposite momenta for the particles and antiparticles forming the pair respectively weakening the inhomogeneity. In conclusion, the magnetic field acting on the particle-antiparticle component of the inhomogeneous phase try to regain the ground state of the magnetic catalysis scenario. In Region 3, with a larger presence of particle-hole pairs, we observe (see Fig. \ref{Fig. 2}-(b)) a milder effect  but favorable of the magnetic field on $m$ and $b$. The shift to the left of the boundary between Region 2 and 3 at a strong magnetic field, as can be seen comparing Fig. \ref{Fig. 1}-(a) and (b), is due to the fact that with the magnetic field the particle-hole density of state increases with the magnetic field  and once the particle-antiparticle pairs' participation is weakened as it is raised the Fermi energy, the presence of the particle-hole pairs is showed  earlier.  

A similar effect for pairs taking place close to the Fermi sphere was found in color superconductivity under a rotated magnetic field where the magnetic moments of the quarks forming the Cooper pairs were also aligned and hence the magnetic field strengthens the couple \cite{CS-Efrain}. In a conventional superconductor, on the contrary, because the particles forming the pair have opposite spins and equal charges, their magnetic moments are pointing in opposite directions and a magnetic field tends to break the pairs by rotating one of the magnetic moments.

The minimum solutions for $m$ and $b$ we are reporting in Figs.\ref{Fig. 1}-\ref{Fig. 2} will be used in the calculations to be performed in the following sections.  
Here, we want to underline that in the calculations done in Figs. \ref{Fig. 1}-\ref{Fig. 2}, as well as in all other numerical calculations done in this paper, the GL expansion (\ref{omega}) is taken up to $N=10$, which, as demonstrated in \cite{Will}, is the order required for the GL expansion to closely approximate the exact thermodynamic potential at $m, b < \mu$ given in Eqs. (70)-(75) of Ref. \cite{Incera}-b.

\section{Permanent Magnetization: Ferromagnetism in dense quark matter}

The magnetization as a function of the magnetic field is given by the formula obtained from (\ref{omega}),
\begin{equation}\label{magnetization}
M^{ ( 2N ) }( B ) = - \frac{ \partial \Omega^{ ( 2N ) } }{ \partial B } = M^{ ( 2N ) }_{ 0 } + \sum_{ n = 1 }^{ \infty } M^{ ( 2N ) }_{ n } B^{ 2 n - 1 }
\end{equation}

Indeed, the series in the magnetic field is an expansion in powers of $B/\mu^2$ and $B/\Lambda^2$, which in the limit we are considering, $B < \mu^2 < \Lambda^2$, which is the suitable one for NS astrophysics, turns appropriate for a weak-field approximation truncation. To see how this happens, let us consider the expansion of the thermodynamic potential (\ref{omega}) up the sixth order \cite{LP-B} as an example. In this case we have,
\begin{eqnarray}\label{GL-Free Energy-qdelta}
\Omega^{(6)}&=&m^2 \left [\alpha_{2,0}+\alpha_{4,2} b^2+\alpha_{6,4} b^4 +\beta_{3,1} b+\beta_{5,3} b^3 \right]+\nonumber
\\
&+&m^4 \left [ \alpha_{4,0}+ \alpha_{6,2} b^2 + \beta_{5,1} b \right ] + m^6 \alpha_{6, 0}.
\end{eqnarray}

Then, the first term in the magnetization (\ref{magnetization}) at order 6 is given by
\begin{equation}\label{magnetization-2}
M_0^{(6)} = m^2\left [\left (\frac{b}{\mu}\right) \bar{\beta}_{3,1}+\left (\frac{m^2}{\mu^2}\right ) \left ( \frac{b}{\mu}\right ) \bar{\beta}_{5,1}+\left (\frac{b^3}{\mu^3}\right )\bar{\beta}_{5,3}\right ]
\end{equation}
where $\bar{\beta}_{i,j}$ are numerical coefficients.

For the rest of the terms in (\ref{magnetization}), we start by taking into account that, as it is known, the first subindexes in the $\alpha$ and $\beta$ coefficients are equal to the sum of the powers of the factors $m$ plus those of $b$; and the second subindexes are equal to the powers of $b$. From the terms with $\beta$ coefficients, it is obtained only linear in $B$ contributions multiplying a series in powers of $(b/\mu)^{2j+1}$ as seen from (\ref{beta}) and (\ref{GL-Free Energy-qdelta}). On the other hand, the $\alpha$ coefficients, $\alpha_{2,0}$, $\alpha_{4,2}$ and $\alpha_{6,4}$, which are the corresponding last members of each family, can be found using formula (\ref{alpha}). The rest of the $\alpha$ coefficients in (\ref{GL-Free Energy-qdelta}) are given by a numerical factor multiplied by the corresponding last member of the family as given in (\ref{generic-coeff}). Hence, to see how the magnetic field enters in each last member of each family, we have to analyze how it enters in the coefficients $\alpha_{2,0}$, $\alpha_{4,2}$ and $\alpha_{6,4}$ of (\ref{GL-Free Energy-qdelta}). Now, from (\ref{alpha}) and (\ref{I-3}), we can see that the $B$ dependence determining their contributions to the  magnetization is given by
\begin{eqnarray}\label{alpha-expansion}
\alpha_{2,0}&=& B \left [C_1\frac{B}{\mu^2}+C_2\frac{B^3}{\mu^6}+C_3 \frac{B^5}{\mu^{10}}+\cdots +(\mu \rightarrow \Lambda) \right ], \nonumber
\\
\alpha_{4,2} b^2&=&  B \frac{b^2}{\mu^2} \left [C'_1\frac{B}{\mu^2}+C'_2\frac{B^3}{\mu^6}+ C'_3\frac{B^5}{\mu^{10}}+\cdots +(\mu \rightarrow \Lambda) \right ], \nonumber
\\
\alpha_{6,4} b^4 &=&  B\frac{b^4}{\mu^4} \left [C''_1\frac{B}{\mu^2}+C''_2\frac{B^3}{\mu^6}+ C''_3\frac{B^5}{\mu^{10}}+\cdots +(\mu \rightarrow \Lambda) \right ] ,
\end{eqnarray}
where $C_i, C'_i$ and $C''_i$ are numerical coefficients.
From (\ref{magnetization-2}) and (\ref{alpha-expansion}), we see that $M^6$ is indeed expressed as an expansion in the small parameters $\frac{b}{\mu}$, $\frac{b}{\Lambda}, \frac{B}{\mu^2}$ and $\frac{B}{\Lambda^2}$ under the approximation $\Lambda^2>\mu^2>b^2>B$.  This fact facilitates the truncation of the series in power of the magnetic field in (\ref{magnetization}).

We should notice moreover that the expansion coefficients $M^{(20)}_i, i=0,2,\cdots$ in (\ref{magnetization}) implicitly depend on $B$ through the condensate parameters $m$ and $b$. But as seen from Figs. \ref{Fig. 1}-(a) and Fig. \ref{Fig. 2}, even at $B=0$ those parameters are different from zero. Therefore, if $M^{(20)}_0\neq0$ when $m$ and $b$ are evaluated at their minimum solutions for  $B=0$, we can say that there exists a permanent magnetization in the system. That is, the system retains a magnetization that prevails even when $B\rightarrow 0$. From now on, to simplify the notation we will skip the supra-index (20) in $M^{(20)}_i$. 

 \begin{figure}
\begin{center}
\begin{tabular}{ccc}
  \includegraphics[width=9cm]{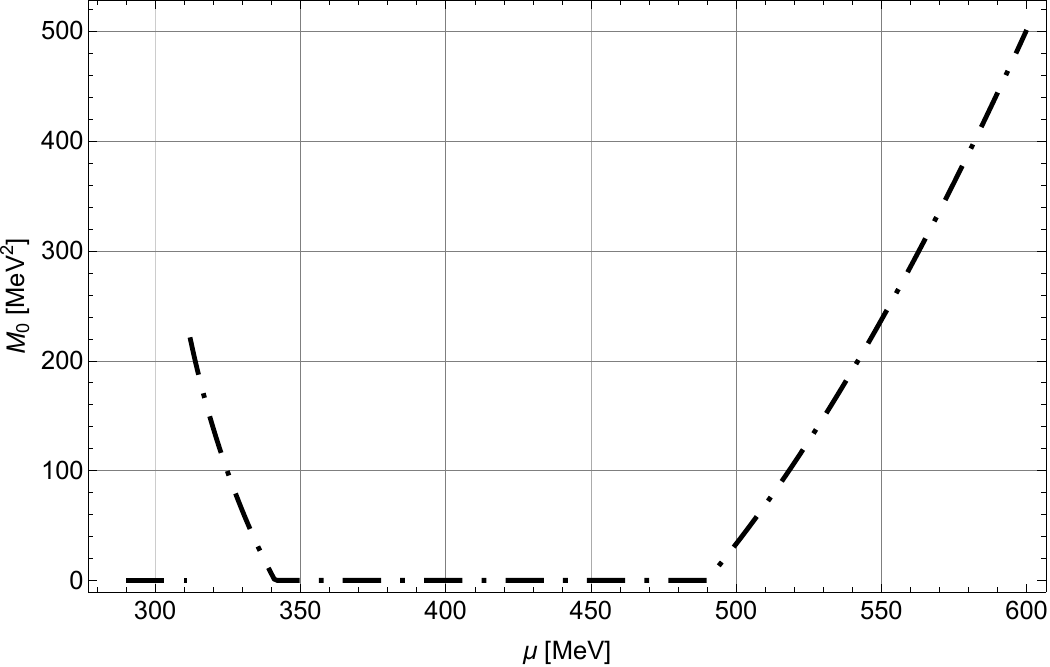} & \includegraphics[width=9cm]{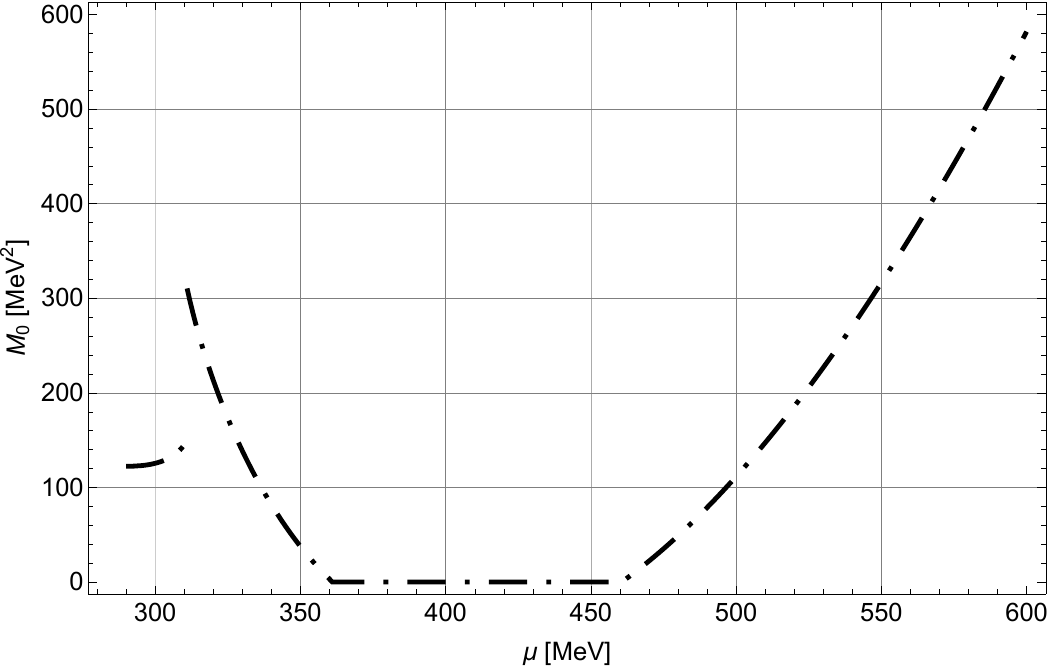}\\
(a) & (b)  \\
  \\
  \end{tabular}
    \end{center}
    \caption{The permanent magnetization, $M_{ 0} (B=0)$, as a function of the chemical potential $\mu$ is given in panel (a), and $M_0(B)$ at $B=1.5 \times10^{18}$ G in (b). Both graphs are obtained from the GL expansion (\ref{omega}) at $N=10$ and with condensate parameters $m$ and $b$ found from the minimum equations (\ref{min-m}) and (\ref{min-b}) respectively.}
     \label{Fig. 3}
\end{figure}

The magnetization $M_0$ is given as a function of the baryonic chemical potential at $B=0$ in Fig.\ref{Fig. 3}-(a) and at $B \neq 0$ in Fig. \ref{Fig. 3}-(b). In doing those plots the parameters $m$ and $b$ are taking as the solutions of Eqs. (\ref{min-m}) and (\ref{min-b}) at $B=0$ and $B \neq 0$ respectively.
 From Fig.\ref{Fig. 1}-(a) and Fig.\ref{Fig. 3}-(a), we see that once the Fermi sphere is formed and the condensate modulation arises, the system acquires a permanent magnetization characterizing a ferromagnetic behavior. Therefore, at a critical value of the baryonic chemical potential $\mu_{0}=312$ MeV there is a discontinuous jump in $M_0$ from zero to a different from zero value. Being $M_0$ a first derivative of the thermodynamic potential with respect to the magnetic field, the jump is signaling a first-order phase transition which is related to the emergence of the inhomogeneous ground state (i.e. with $b \neq 0$).
 
 \begin{figure}
\begin{center}
\begin{tabular}{ccc}
  \includegraphics[width=9cm]{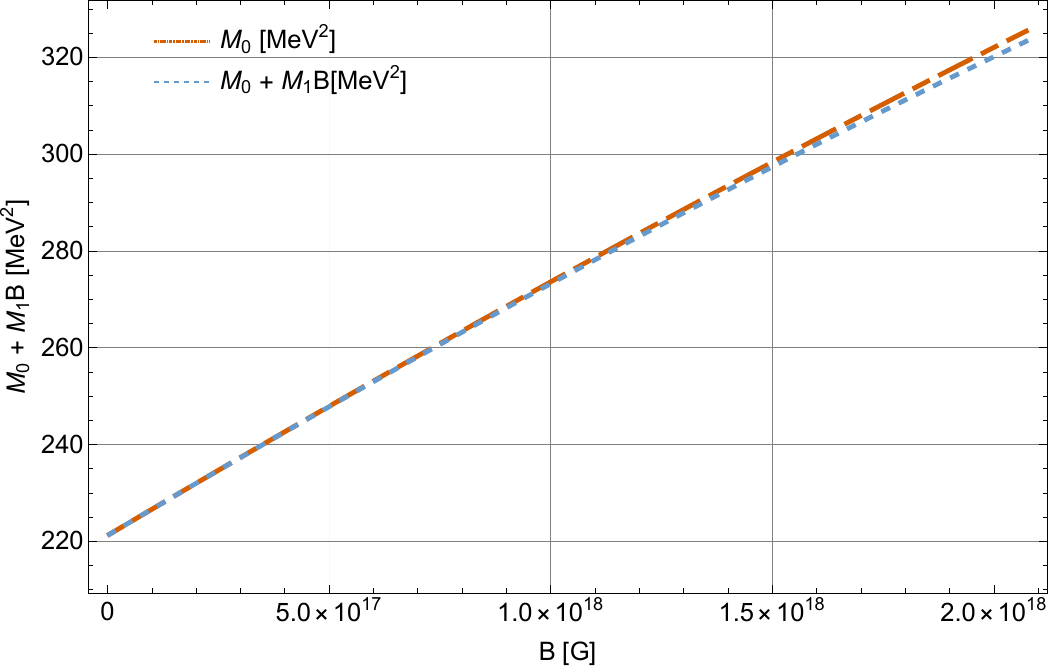} & \includegraphics[width=9cm]{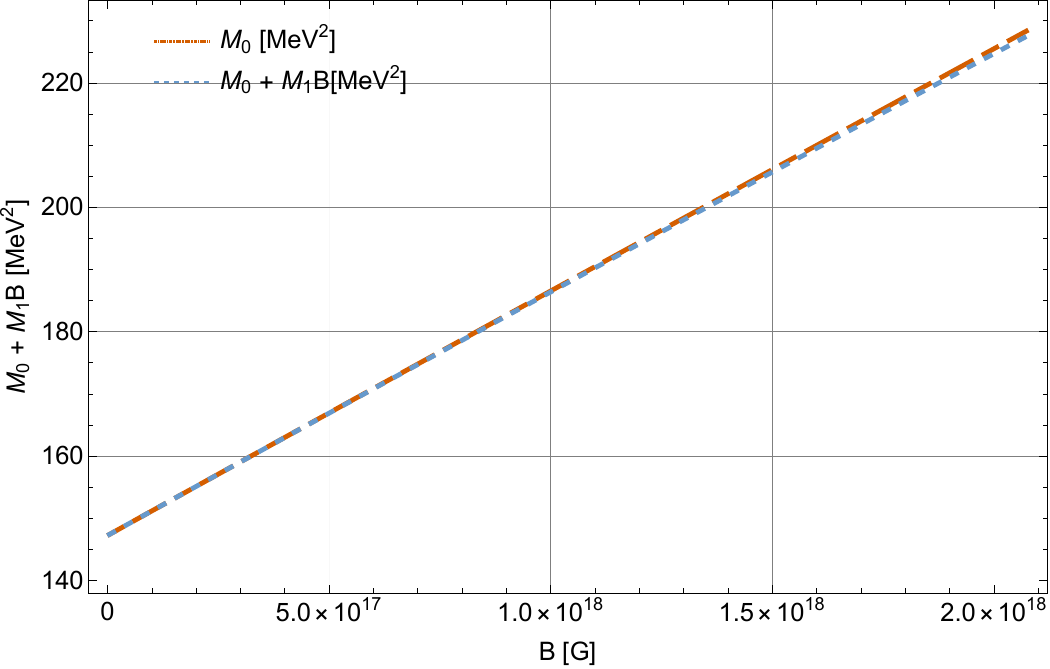}\\
(a) & (b)  \\
  \\
  \end{tabular}
    \end{center}
    \caption{(Color online) Magnetization $M(B)$ as a function of magnetic field at $\mu=312$ MeV in (a) and at $\mu=530$ MeV in (b). In each graph we plotted $M_0$ and $M_0+M_1 B$ versus $B$.}
     \label{Fig. 4}
\end{figure}

The result reported in Fig.\ref{Fig. 3}-(a) up to $\mu=342$ MeV coincides, apart from some numerical factors, with the one found in Ref. \cite{Tatsumi-M0} that was obtained in the small $m$ limit from the exact thermodynamic potential. The numerical discrepancy between the two results is due to the fact that we are including the color number $N_C=3$ and considering the values for the NJL model parameters $\Lambda$ and $G$, used in Ref.  \cite{KlimenkoPRD82}, which differ from those used in \cite{Tatsumi-M0}. 

To avoid the possible suspicion of a non analytic behavior of $M(B)$ at $B \rightarrow 0$, we are showing in Fig.\ref{Fig. 4} the graphs for $M(B)$ vs $B$ at  $\mu=312$ MeV in panel (a) and at $\mu=530$ MeV in panel (b), which correspond to chemical potentials in Regions 1 and 3 of Fig. \ref{Fig. 1} respectively. In both plots we can corroborate that the magnetization smoothly tends to $M_0(B=0)$ when $B \rightarrow 0$, so showing that this limit is analytic. Thus, this result is indicative that there exists a permanent magnetization in the system when the magnetic field tends to zero.

Here, the following discussion is in order. The existence of the found permanent magnetization $M_0(B=0)$ is related to the anomalous effect produced by the quarks in the LLL. Nevertheless, it could seem that there is a contradiction by the fact that $M_0\neq0$ when evaluated at $B=0$. But, we should notice in this regard that the contribution to $M_0$ from the thermodynamic potential (\ref{omega}) comes from the $\beta$-coefficient terms, which, as we already pointed out, are directly related to the LLL contribution to $\Omega^{(2N)}$. Thus, what is determining the permanent magnetization is the structure of the LLL contribution to the thermodynamic potential $\Omega^{(2N)}\approx M_0 B$, which still remains in the $B\rightarrow 0$ limit of $M_0$, since $m$ and $b$ are different from zero in the $B \rightarrow 0$ limit as seen from Fig.\ref{Fig. 2}. From a physical point of view, we have that in the presence of a magnetic field the quarks (anti-quarks) occupying the LLL have only one spin orientation (with the anti-quarks having the opposite orientation). Thus, the magnetic field orients the corresponding magnetic moment so to minimize the system energy (the magnetic moment of the anti-quarks will have the same orientation since they have opposite spin and opposite charge). Disconnecting the magnetic field, as suggested in Fig. \ref{Fig. 4} by taking the $B \rightarrow 0$ limit, the magnetic moment of the pair that minimized the energy is preserved so keeping the system magnetization $M_0(B=0)$,  similarly to what happens with ferromagnetic materials that acquire a permanent magnetization once the magnetic field is turned off. Hence, we can see the important role that the LLL is playing in this phenomenology. We must also emphasize that not every system under a magnetic field can exhibit this effect due to the contribution of LLL.  We should underline, that in the MDCDW case the joint effect of the existence of an inhomogeneous condensate that breaks time-reversal symmetry, thus allowing a term $M_0 B$ in the free energy, and the asymmetry of the quark spectrum in the LLL, which determines the analytical structure of $M_0$,  played a crucial role in the realization of this anomalous magnetization effect.

Finally, to facilitate a more solid explanation of the paradoxical physical picture under discussion, we must point out that in this dense quark phase there will always be present a magnetic field spontaneously induced once $m$ and $b$ are different from zero, as we will see in the next section. Thus, $M_0$ will always depends on the magnetic field through $m$ and $b$ in this phase.

\begin{figure}
\begin{center}
\begin{tabular}{ccc}
  \includegraphics[width=7cm]{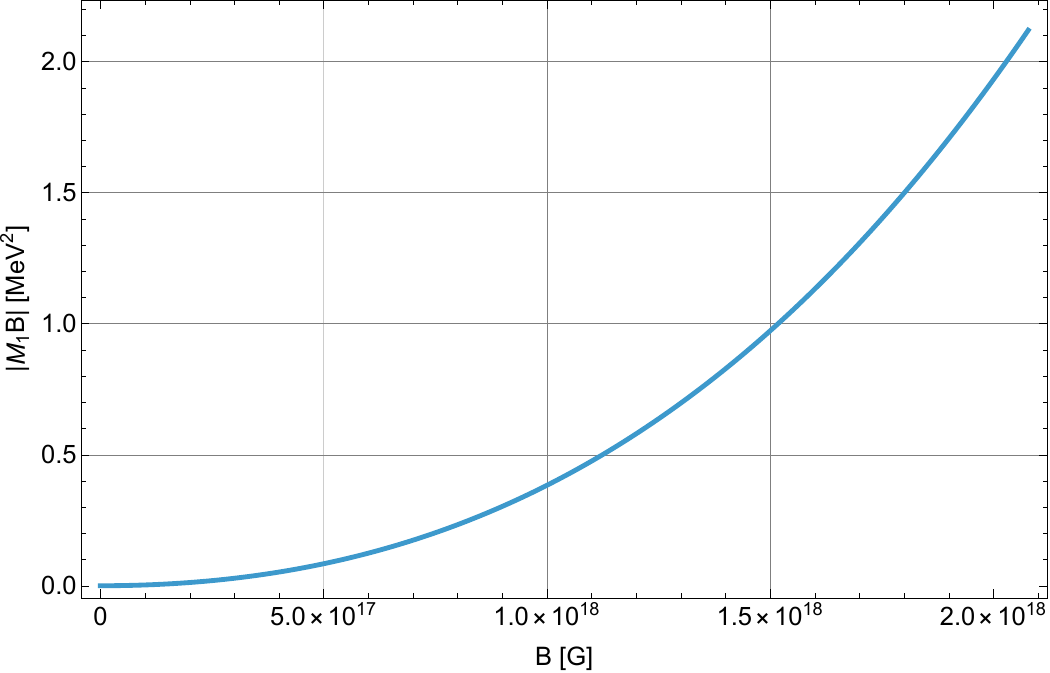} & \includegraphics[width=7cm]{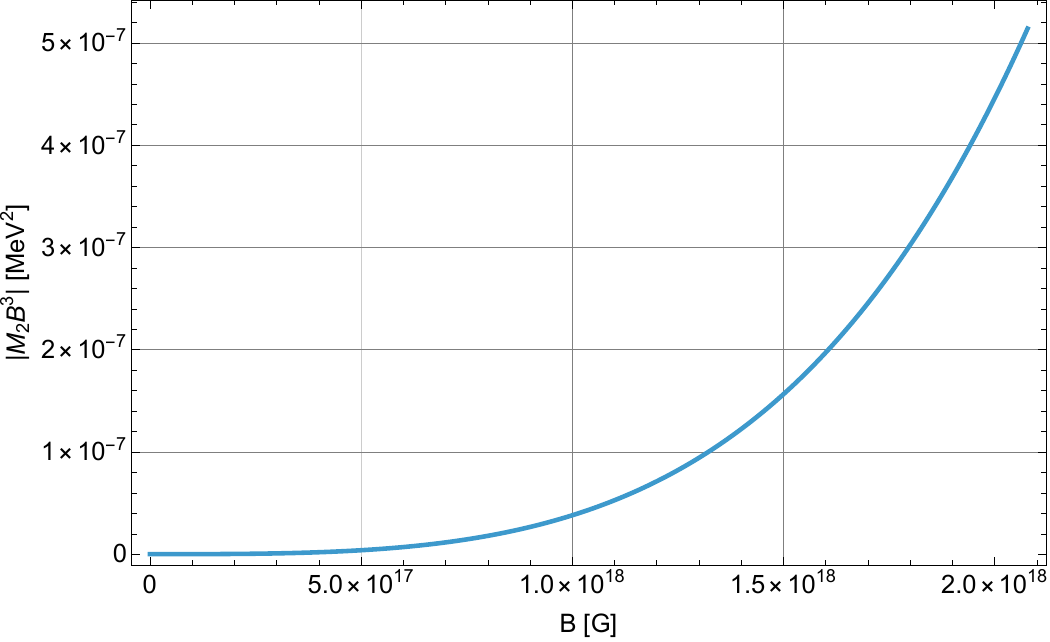}\\
(a) & (b)  \\
  \\
  \end{tabular}
    \end{center}
    \caption{Contributions of $M_1B$ in (a) and $M_2 B^3$ in (b) versus magnetic field for $\mu=312$ MeV}
     \label{Fig. 5}
\end{figure}

To determine where to cut the expansion in powers of $B$ in (\ref{magnetization}), we plot in Fig. \ref{Fig. 5}-(a) and (b) the magnetization contributions linear and cubic in $B$  versus the magnetic field respectively. Comparing the values of those terms with the ones of Fig. \ref{Fig. 4}, we see that the contribution of terms beyond $M_0$ is negligibly small. Since the terms $M_i$ with $i\geqslant 1$ get only contributions from the HLL, while $M_0$ only depends on the LLL, we conclude that the magnetization of this system is mainly driven by the dynamics of the LLL. In Fig. \ref{Fig. 4} we showed that adding the term $M_1B$ to $M_0$ does not make a significant change in the magnetization versus magnetic field curves. 

\section{Induced magnetic field in the DCDW phase of dense quark matter}

The expansion of the thermodynamic potential up to second order in powers of the magnetic field is given by
\begin{equation}\label{Eff-Potential}
\Omega(B, E) \simeq \frac{ 1 }{ 2 } \left( B^{ 2 } \right) - M_{ 0 } B ,
\end{equation}
where the first term in the right-hand side is the Maxwell term, and the second term is the medium term obtained from (\ref{omega}) to order $N=10$.  Our goal now is to investigate the possibility to have an induced magnetic field in the DCDW phase where the inhomogeneous condensate (\ref{DCDW-cond}) has been spontaneously generated. From a symmetry point of view, since the DCDW inhomogeneous condensate spontaneously breaks time-reversal and rotation $O(3)$ symmetries, an induced magnetic field does not break any symmetry that has not already been broken by the condensate forming the ground state, so lacking of a mechanism that protect the system against the spontaneous induction of a magnetic field.

\begin{figure}[H]
    \centering
    \includegraphics[width=0.7\linewidth]{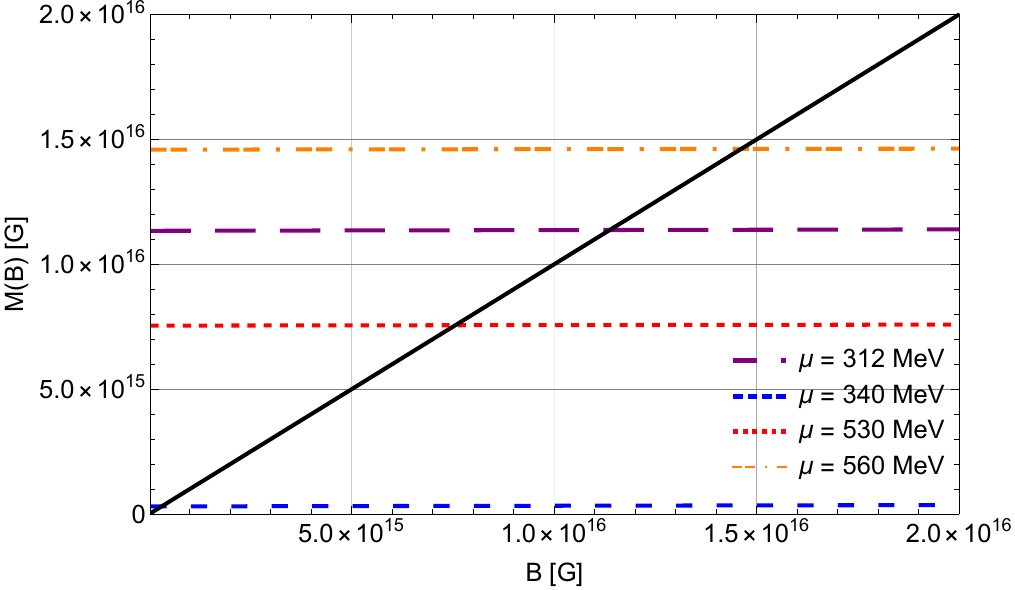}
    \caption{(Color online) Graphical method to get the solutions of the induced magnetic field for different chemical potentials. The found solutions are respectively, ( $\mu=312$ MeV , $B=1.19  \times 10^{16}$ G), 
  ($\mu=340$ MeV,  $B=3.19\times10^{14}$ G), ($\mu=530$ MeV, $B=7.5 \times 10^{15}$ G), ($\mu=560$ MeV, $B=1.45\times10^{16}$).G}
    \label{Fig_6}
\end{figure}

Since, $M_0$ implicitly depends on $B$, to solve the minimum equation for $B$
\begin{equation}\label{Minimum-Eq}
\frac{ \partial \Omega }{ \partial B } = B-M_0=0, 
\end{equation}
we use the graphical method shown in Fig. \ref{Fig_6}, where the dashed lines correspond to $M_0(B)$ versus $B$ for different chemicals potentials and the continue line representing the $B=B$ graph, which is drown at $45^0$. The interception point for each  $M_0(B)$  gives the solution of  (\ref{Minimum-Eq}) for the corresponding chemical potential. We selected two chemical potentials corresponding to Region 1 and two other corresponding to Region 3 and the corresponding values of the induced magnetic field are for $\mu=312$ MeV, $B=1.19  \times 10^{16}$ G, for $\mu=340$ MeV,  $B=3.19\times10^{14}$ G, for $\mu=530$ MeV, $B=7.5 \times 10^{15}$ G, and for $\mu=560$ MeV, $B=1.45\times10^{16}$ G.

From this set of values we observe that when increasing the chemical potential in Region 1, the magnitud of the induced magnetic field decreases, at the same time that the amplitude of the condensate decreases; while in Region 3 when the chemical potential increases, the condensate magnitude increases as well as the value of the corresponding induced magnetic field. Hence, we see that there is a direct proportionality between the magnitude of the induced magnetic field and the magnitude of the condensate in both regions. Since the magnitude of the condensate is related to the pair condensation energy we conclude that the energy needed to create a magnetic field is supplied by the lowering of the system energy due to the condensation process.

\section{Induced elecric field in the DCDW phase of dense quark matter}

As it was explained in Section 2, in the presence of collinear electric and magnetic fields, a new term is raised in the regularized effective action of the MDCDW phase \cite{Incera}.
The  new contribution in the regularized effective action turned out to be the chiral anomaly in the electromagnetic
sector $(\kappa/8)\theta F^{\ast}_{\mu \nu}F^{\mu\nu}$, with $\kappa/8=\alpha/4\pi$. Thus, the interaction between the electromagnetic strength tensor and its dual are coupled through $\theta = 2bz$, with $z$
the spatial coordinate in the direction of the condensate modulation.

Thus, the expansion of the thermodynamic potential up to second order in powers of the fields is given by
\begin{equation}\label{Eff-Potential}
\Omega(B, E) \simeq \frac{ 1 }{ 2 } \left( E^{ 2 } + B^{ 2 } \right) - \frac{ \kappa }{ 2 } \theta \vec{ B } \cdot \vec{ E } - M_{ 0 } B .
\end{equation}

To have a term linear in $B$ in $\Omega(B, E)$, the system ground state should  break the time reversal symmetry, which is the case in the MDCDW phase under study  \cite{fMagnetoelectricity}.
In addition, to have such a term, we need covariant scalar structures that allow for such linear term in the field. For the magnetization times magnetic field term, we can see that the thermodynamic potential in this system can have a structure proportional to $\epsilon_{\mu\nu\rho\lambda}u^\mu n^\nu F^{\rho\lambda}$, where $u^\mu$ is the medium four-velocity, which in the rest frame takes the form $u_\mu=(1,0,0,0)$, and  $n^\nu$ is a four vector associated with the modulation of the inhomogeneous condensate, which in the rest frame takes the form $n_\nu=(0,0,0,1)$. The magnetic field along the z-direction is given by the tensor $F_{21}$. But, to have an electric polarization term entering in $\Omega(B, E)$ as $P_0 E$, parity symmetry should be broken, which is not the case in the system under consideration. 
Moreover, to sustain such term there is no tensor covariant structure that can represent a term linear in the electric field of the form $\epsilon_{\mu\nu\rho\lambda}u^\mu n^\nu F^{\rho\lambda}$. Thus, 
the only term linear in the electric field which preserves parity symmetry is of the form $(\kappa/8)\theta\epsilon_{\mu\nu\rho\lambda}F^{\mu\nu}F^{\rho\lambda}$, which is the anomalous second term in the right-hand-side of (\ref{Eff-Potential}).

Now, to find if the $B$ and $E$ can be spontaneously induced, we need to consider the minimum equations
\begin{equation}\label{Minimum-Eq}
\frac{ \partial \Omega }{ \partial B } = 0, \quad  \frac{ \partial \Omega }{ \partial E } = 0,
\end{equation}
which give rise respectively to the couple equations

\begin{equation}\label{Minimum-B}
\frac{ \partial \Omega }{ \partial B } = B - M_{ 0 } - \frac{ \kappa }{ 2 } \theta E = 0,
\end{equation}

\begin{equation}\label{Minimum-E}
\frac{ \partial \Omega }{ \partial E } = E - \frac{ \kappa }{ 2 } \theta B = 0,
\end{equation}
where we are considering that both $B$ and $E$ are along the z direction and we are neglecting $M_i,  i\geqslant 0$ since they are too much smaller than $M_0$, as seen from Fig. \ref{Fig. 5} and discussed previously.

Substituting with (\ref{Minimum-E}) in (\ref{Minimum-B}) we obtain

\begin{equation}\label{Minimum-B-1}
\frac{ \partial \Omega }{ \partial B } = B - M_{ 0 } - \frac{ \kappa^{ 2 } }{ 4 } \theta^{ 2 } B = \left( 1 - \frac{ \kappa^{ 2 } }{ 4 } \theta^{ 2 } \right) B - M_{ 0 } = 0.
\end{equation}

Taking into account that $\theta=2bz$, we will consider a domain of the order of the wave length of the inhomogeneous condensate $\lambda_b=1/2b$. Then, the average value of $B$ in that domain is,

\begin{equation}\label{Solution-B}
B = \frac{ 1 }{ \lambda_{ b } } \int _{ 0 }^{ \lambda_{ b } } \dd{ z } \frac{ M_{ 0 } }{ \left( 1 - ( b \kappa z )^{ 2 } \right) } = \frac{ 2 M_{ 0 } }{ \kappa } \int_{ 0 }^{ \frac{ \kappa }{ 2 } } \frac{ \dd{ z' } }{ 1 - z'^{ 2 } }\approx M_0,
\end{equation}

and from (\ref{Minimum-E})  we have,

\begin{equation}\label{Solution-E}
 E = \frac{ 1 }{ \lambda_{ b } } \int _{ 0 }^{ \lambda_{ b } } \dd{ z } \frac{ M_{ 0 } \kappa b z }{ \left( 1 - ( b \kappa z )^{ 2 } \right) } = 2 M_{ 0 } \int_{ 0 }^{ \frac{ \kappa }{ 2 } } \frac{ \dd{ z' } }{ 1 - z'^{ 2 } } = \kappa B,
\end{equation}
where we did the variable change $z'=b \kappa z$. 

We have that Eq. (\ref{Solution-B}) coincides with (\ref{Minimum-B}), which was already solved numerically in Section 4.
On the other hand, from (\ref{Solution-E}) we see that $E \ll B$, what is expected from the fact that $E$ has a quantum origen of order $\alpha$, which is produced through the anomaly by radiative effects that can take place once a magnetic field is generated in the system. Hence, we have that $| E|  \approx 10^{ - 3 } |B|$.

The values of $B$ and $E$ depend on $\mu$ through $M_0$. For example, in the mid-density region (Region 1: from $\mu=312$ MeV to $\mu=342$ MeV) the induced magnetic field, for instance, can go from zero to $10^{16}$ G, with $E$ damped in three orders.

The spontaneous induction of the permanent magnetization $M_0$, together with a magnetic and an electric field have significant consequences. It implies that the transit between the two phases $DCDW \rightarrow MDCDW$ occurs spontaneously once the ground state is realized by the condensate (\ref{DCDW-cond}) that breaks time-reversal and rotational symmetries as needed for the presence of the induced $E$ and $B$ fields respectively. This transit, on the other hand, removes the LP instability, which could otherwise disable the realization of this single-modulated DCDW phase. All these anomalous behaviors can be traced back to the asymmetry with respect to the zero energy level in the spectrum of the LLL (\ref{LLL}), which is created once a magnetic field is induced on that ground state. Thus, dense quark matter in this phase exhibits an intrinsic mechanism for generating electromagnetic fields, which can be of great interest for magnetar physics. 


\section{Concluding remarks}

The effect of a magnetic field on the chiral condensate was exhaustively studied at $\mu = 0$ and neglecting the field effect on the pairing coupling in Refs. \cite{Magnetic Catalysis}. There, it was found that the magnetic field strengthen the chiral condensation in the weak coupling limit. That phenomenon was known as the Magnetic Catalysis of Chiral Symmetry Breaking.  Later on, adding the effect of the magnetic field on the pairing coupling it was found that the magnetic field can weaken the chiral condensation \cite{Inverse Mag. catalysis}. This phenomenon is called the Inverse Magnetic Catalysis. When the baryonic chemical potential is added as a new factor, it was first found that at a critical value $\mu_c \gtrsim m_{dyn}$, where $m_{dyn}$ is the dynamical mass due to the chiral condensate at $\mu=0$, the constant chiral condensate disappears \cite{MCSB-Density}. However, it was shown later that at $\mu \neq 0$ the phase with the inhomogeneous condensate  (\ref{DCDW-cond}) is energetically preferred over the homogeneous chiral phases (broken and unbroken) in a wide range of the parameter space at supercritical coupling and even at subcritical coupling \cite{Hidaka, DCDW, Bo, KlimenkoPRD82}. Also it was found that the condensate under sufficiently strong
magnetic fields, is favored up to temperatures of tens of MeV, which is several orders of magnitude higher than typical temperatures of old NSs \cite{Will-2, Will}. 

The inhomogeneous chiral condensate (\ref{DCDW-cond}) is realized once $\mu \gtrsim m_{dyn}$ even at $B=0$ forming what is known as the DCDW phase \cite{DCDW}. Nevertheless, single-modulated condensates, as the one of the DCDW phase (\ref{DCDW-cond}), suffer from the LP instability \cite{Landau} that takes place due to the thermal fluctuations at any nonzero temperature. 
Consequently, the LP instability undermines the long-range order of single-modulated inhomogeneous phases in 3+1
dimensions. However, as it was shown in \cite{Will-2, LP-B}, the situation changes for the DCDW phase in the presence of a magnetic field. 
Indeed, the joint effects of the rotational symmetry breaking produced by the presence of the magnetic field together with the fact that the inhomogeneous condensate (\ref{DCDW-cond}) breaks time-reversal symmetry open the possibility for free-energy terms which are odd in the condensate modulation, to exist.  These odd-in-$q$ terms ensure the absence of soft transverse modes in the spectrum of the fluctuations, hence the lack of LP instability \cite{Will-2, LP-B}.

With the results we are reporting in this paper we are showing that once the Fermi sphere is formed at moderate baryonic density and 
the ground state of quark matter takes the inhomogeneous form  (\ref{DCDW-cond}) the system spontaneously undergoes the phase transition DCDW $\rightarrow$ MDCDW with the induction of a magnetic and a much weaker electric fields.

Under these conditions, the system is spontaneously released from the thermal  LP instability. Once the pair condenses, spontaneously selecting a modulation direction, the induced magnetic field will arise pointing  in the same direction as the one of the condensate modulation, since the alignment of both is energetically preferred  \cite{KlimenkoPRD82, PLB743}. Moreover, because of the chiral anomaly term, an electric field will be induced, also collinear to the modulation direction, but with a magnitude three orders lower than that of the magnetic field. Thus, our result is proposing a new mechanism to generate magnetic and electric fields in NSs endowed with dense quark matter in their interiors, without the necessity of any magneto-hydrodynamic mechanism.

Let us discuss a possible mechanism that can serve to create and propel inner magnetic fields in magnetars. For this analysis we will restrict ourselves to Region 1 and assume that the inner density of NS is in that range of densities, which is not unreasonable. The arguments can be also extended to Region 3.

From the values of the induced magnetic fields obtained from Fig. \ref{Fig_6}, we see that as the density approaches  $\mu\simeq 340$ MeV the value of the induced inner magnetic field will be smaller. 
But if it turns out that the MDCDW phase could coexist with the 2SC color-superconducting phase, then we would have two complementary mechanisms one due to the inhomogeneous chiral condensate that can generate a magnetic field, does not matter how weak, and the other, produced by the color superconductor ground state in the presence of a magnetic field that will boost the magnetic field to the expected inner values  for NSs. In this case, the inhomogeneous chiral condensate would serve to generate the magnetic field that, in the color-superconducting phase, would be amplified by the zero-mode mechanism \cite{Bo-Efrain}.

The zero-mode mechanism refers to the generation of a tachyonic mode as a consequence of the interaction of a magnetic field with the anomalous magnetic moment of spin-1 charged fields. This instability occurs when the field becomes greater than the square of the Meissner mass of the spin-1 fields, and the solution is given by the restructuring of the ground state of the system, giving rise to vortex solutions that, as a result, produce a boosting of the magnetic field. This effect has been reported in different systems of spin-1 charged fields \cite{Zero Mode} and specifically in color superconductivity where certain gluons acquire a rotated charge \cite{Zero Mode-2}. The fact that at moderate densities the 2SC color superconducting phase shows the so-called chromomagnetic instability contributes substantially to the possibility of this scenario being realized \cite{Bo-Efrain}. In this regard it is still an open question if the MDCDW phase and the magnetized 2SC phase of color superconductivity can coexist at moderate densities.

{\bf Acknowledgments:} We thank William  Gyory and Vivian de la Incera for valuable
discussions and comments. The work of E. J. F. was supported in part by National Science Foundation Grant No. PHY-2013222 and Department of Energy Grant No. DE-SC0022023, the work of  J. M. P-F was supported by Department of Energy Grant No. DE-SC0022023.

\end{document}